\documentstyle[epsfig,sprocl]{article}

\def\Journal#1#2#3#4{{#1} {\bf #2}, #3 (#4)}

\def\NPB{{\em Nucl. Phys.} B}
\def\PLB{{\em Phys. Lett.}  B}
\def\PRL{\em Phys. Rev. Lett.} 
\def\PRD{{\em Phys. Rev.} D}
\def\ZPC{{\em Z. Phys.} C}

\def\be{\begin{equation}}
\def\ee{\end{equation}}
\def\bea{\begin{eqnarray}}
\def\eea{\end{eqnarray}}

\newcommand{\mIm}{\,\mbox{$\Im$m\,}}
\newcommand{\eRe}{\,\mbox{$\Re$e\,}}
\newcommand{\n}{\noindent}

\def\gsim{\lower0.5ex\hbox{$\:\buildrel >\over\sim\:$}} 
\def\lsim{\lower0.5ex\hbox{$\:\buildrel <\over\sim\:$}}
\def \cp{{\rm CP}\!\!\!\!\!\!\!\!\!\not~~~~}
\def \cpint{{\cp\mid_t}}
\def \cpinsub{CP$\!\!\!\!\!\!\!\not~~~~~$}

\begin{document}

\title{CP-VIOLATION IN THE TOP SECTOR
\footnote{Talks presented at the Joint Minerva-GIF Symposium, Jerusalem, 
Israel, May 1998 (SBS) and at the Workshop on CP Violation,
Adelaide, Australia, July 1998 (GE).}
}

\author{S. BAR-SHALOM}

\address{Department of Physics, University of California,
Riverside,\\ CA 92521, USA\\E-mail: shaouly@phyun0.ucr.edu} 

\author{G. EILAM}

\address{Department of Physics, Technion-Israel Institute of Technology,\\
Haifa 32000, Israel\\E-mail: eilam@physics.technion.ac.il}

\maketitle\abstracts{We discuss the prospects--within several models--for 
the observation of CP-violation ($\cp$) in top decays and production.
The outlook looks best for $t \to bW$ at the LHC (MSSM $\cp$), $t \to
b \tau \nu_\tau$ at TeV3, LHC and NLC ($H^+$ $\cp$), $p \bar p \to
t \bar b + X$ at TeV3 (MSSM $\cp$), $p p \to t \bar t + X$ at the LHC 
(MSSM $\cp$ \& neutral Higgs $\cp$) and for 
$e^+ e^- \to t \bar t h,~t \bar t Z$, where $h$ is the 
lowest mass neutral Higgs boson,
at an NLC with energy $\geq 1$ TeV (neutral Higgs $\cp$.)}

\section{Introduction or\,: Why Do $\cp$ in the Top Sector?}
Just one experimental number for $Br(K_L\to2\pi)$,~\cite{cronin64}
and a few more that followed, but all in the neutral $K$ system,
induced much activity. One can describe in broad terms most
of the work as trying to answer the following questions:

\begin{itemize}
\item Regarding the $K$ sector: Is mixing  
the only source of the small $\cp$ observed?
\item For the $B$ sector: Is the SM  
the only source of the large $\cp$
expected?  
\item In the $t$ sector: What extension of 
the SM causes the typically small $\cp$
expected and in what observables is it best manifested?
\end{itemize}
      
\n Trying to answer the last question, 
there are already ${\cal O}(100)$ papers on 
$\cp$ for the top sector.~\cite{ourreview}

The advantages for studying $\cp$ in t ($\cpint$) are:

\begin{enumerate}
\item The SM causes vanishingly small 
$\cpint$.~\cite{ehs91} This is due to its very small mixing with
the other generations and to the fact that
$m_t >> m_{\rm other \,quarks}$, 
causing extremely effective
GIM~\cite{gim} cancellations.
\item The top at $\sim 175$ GeV is--among the so far observed 
particles--the closest
to New Physics which may well be the cause for
$\cp$. Indeed, $m_t$ drives large $\cp$
in extensions of the SM. 
\item
If $m_t > m_{\rm new~particles}$, where this ``window'' may close pretty soon,
$\cp$-odd $T_N$-even observables,\footnote{
The ``naive'' time reversal operator $T_N$, operates the same
way as $T$ does, but without flipping between initial and final states.}
which require
absorptive cuts, may be measurable.
\item Since the top is that heavy, it decays before 
it has time to bind into 
hadrons.~\cite {bigiplb86} $\cpint$ is
therefore direct, there are no hadronic complications 
and the information  carried by top spins is not diluted
by hadronization.
${\vec s}_t$ is then as good as any four momentum $p$.
\item Last, but not least, future colliders will 
provide a large number of tops. In fact, the expected number of
$t \bar t$ events/year is $10^4 -10^5$ at the
Tevatron runs $2 - 3$, 
$10^7 -10^8$ at the LHC
and $10^5 -10^6$ at an NLC.
\end{enumerate}

Convinced of the importance of $\cpint$, we proceed as follows:
In Sec. 2 we present models with non-standard $\cp$.
$\cp$ in top decays and CP-violating top dipole moments  
are discussed in Secs. 3 and 4, respectively.
In Sec. 5 we discuss $\cp$ in future collider experiments resulting from:
$f \bar f \to t \bar t$ at TEV3, LHC and NLC,
$u \bar d \to t \bar b$ at TeV3, and
$e^+e^- ~ \to t \bar t h,~t \bar t Z$ (tree-level $\cp$) at an NLC.
Finally, Sec. 6 includes summary and outlook.

\section{Models With Non-Standard $\cp$ (MHDM\protect\footnote
{*HDM, with *=2, 3, or Multi, denotes * Higgs Doublet Models.}  
and Supersymmetry)}

\subsection{\cpinsub in the Neutral Higgs Sector in MHDM} 
Generically, $\cp$ in the neutral Higgs sector of any MHDM,
can be expressed by the presence of both scalar and
pseudoscalar couplings to fermions:

\be
{\cal L}_{h ff}=-\frac{g_W}{\sqrt 2} 
\frac{m_f}{m_W} h \bar f \left( a_f + i b_f \gamma_5 \right) f,\label{eq:hff}
\ee

\n therefore, $\cp$ in $h t \bar t$ is proprtional to $m_t b_t$.
In type II 2HDM 
$b_t$ is proportional to $1/\tan\beta$, where $\tan\beta \equiv v_u/v_d$
($(v_u,~v_d$ are the vacuum expectation values of $H_u,~H_d$, respectively;
$H_q$ endows quark $q$ with a mass).
Note that such a coupling does not exist in the 
Minimal SuperSymmetric Model (MSSM). 

\subsection{\cpinsub in the Charged Higgs Sector in MHDM} 
A generic parameterization for any MHDM is (for each generation):

\be
{\cal L}_{H^+ud}= 
\frac{g_W}{\sqrt{2} m_W} V_{ud}^{\rm CKM} \times 
H^+\bar u (m_u U_u {\bf P}_L + m_d U_d{\bf P}_R) d,\label{eq:Hplusff}
\ee

\n where ${\bf P}_L~,{\bf P}_R \equiv (1 \mp \gamma_5)/2$, 
and a similar equation for ${\cal L}_{H^+ \nu_\ell \ell}$
with $u \to \nu_\ell$ and $d \to \ell$.
$U_u,~U_d,~U_\ell$ depend on the mixing
parameters in the charged Higgs sector, and are
in general complex with at least 3 Higgs doublets.
$H^+$ $\cpint$ is then proportional
to $\mIm(U_u^*U_d)$, or to $\mIm(U_u^*U_\ell)$.

\subsection{\cpinsub in the MSSM}\label{subsec:cpinmssm}
$\cp$ can originate in the MSSM from various sources, as discussed 
below.  Again, the top sector is the best one for studying such effects. 
~\\
\underline{$\cp$ in ${\tilde f}_L -{\tilde f}_R$ 
mixing:} \\
One can parametrize ${\tilde f}_L -{\tilde f}_R$ mixing as follows:

\bea 
{\tilde f}_L &=& \cos\alpha_f {\tilde f}_1 - 
e^{i \beta_f} \sin\alpha_f {\tilde f}_2 \nonumber \\
{\tilde f}_R &=& e^{-i \beta_f} \sin\alpha_f {\tilde f}_1 
+ \cos\alpha_f {\tilde f}_2, 
\eea

\n causing $\cp$ to be proportional to $\xi_{CP}^f 
\equiv \frac{1}{2} 
\sin(2\alpha_f) \sin\beta_f$. Now, since 
$\sin(2\alpha_f)$ goes as 
$(m_f/M_{\rm Supersymmetry})$, this mechanism is most useful for $f =  t$.
Furthermore, $\beta_t$ is proportional to ${\rm arg}(A_t$ is the
coefficient of $Q H_2 U$ in ${\cal L}_{\rm soft}$).
${\rm arg}(A_t)$ is expected to be small in
the case of complete universality of the soft breaking
parameters at the GUT scale, i.e., N=1 minimal 
supergravity.~\cite{garistoprd97}
However, ${\rm arg}(A_t) \sim {\cal O}(1)$ is consistent with non-universal
$A$ terms. Consequently, we can have  maximal $\cp$ {\it i.e.,} 
$\xi_{CP}^t =\frac{1}{2}$.~\cite{basprd98} 
~\\
\underline{$\cp$ via ${\rm arg}(\mu)$:} \\
One can, in principle, have a non-vanishing ${\rm arg}(\mu)= 
-{\rm arg}(B)$ ($\mu B$ is  the coefficient of $H_1 H_2$ in 
${\cal L}_{\rm soft}$).
However, since $d_n < 0.97 \times 10^{-25}$ e-cm,~\cite{dninpdg}
where $d_n$ is the neutron Electric Dipole Moment (EDM),
${\rm arg}(\mu) \lsim 
10^{-2}-10^{-3}$ is practically inevitable
for $m_{\rm squark} \lsim 1$ TeV.~\cite{9607269}

\section{$\cp$ in Top Decays}
Let us now discuss $\cp$
for top decays to $d_k W$ and 
in particular to $b \tau \nu$.
\subsection{Partial Rate asymmetry (PRA)}
If a PRA is $\neq 0$ then CP is violated.
Since from CPT invariance 
$\Gamma_{\rm total}^{t} = \Gamma_{\rm total}^{\bar t}$, 
once a PRA is non-vanishing,
there should be a compensating PRA from another available 
decay channel. This CP-CPT connection generally dictates 
a small PRA$\mid_t$.
~\\
\underline{PRA in the SM:} \\
${\rm PRA}(t \to d_k W)$ is proportional to
$\mIm{\rm (self-energy-like)} \times\eRe{\rm (tree)}$.
At best it is around $10^{-9}$ for the small $t \to dW$ $(k=1)$ process.
~\cite{mabrandzp92,gkplb93}
${\rm PRA}(t \to u  d_k {\bar d}_k,~c d_k {\bar d}_k)$, 
results from  $\mIm{\rm (``tree'')}$ 
$\times\eRe{\rm (strong~penguin)}$ and
despite $W$ resonance enhancement in the $W$ 
exchange ``tree'' diagram, is $\sim 10^{-5}$
at best for the small $t \to dc\bar d$ decay.~\cite{ehs91}
Therefore, within the SM, $\cpint$ is too small to be measurable.        
~\\  
\underline{PRA in 2HDM:}\\ 
For ${\rm PRA}(t \to d_k W)$, $H^+$ was added to SM $W^+$;~\cite{gkplb93} 
no enhancement over the SM was found. Furthermore, there is
no contribution from the neutral Higgs $h$ in 
2HDM with natural flavor conservation, {\it e.g.,} type II.
~\\
\underline{PRA in 3HDM:}\\ 
${\rm PRA}(t \to b \tau \nu_\tau)$ goes like 
$\mIm{\rm (W-``tree'')} \times\eRe{\rm (H-tree)}$ and is 
proportional to $\mIm G^W_L \lsim 10^{-8}$,~\cite{sehgalprd}
where $G_L$ is the longitudinal part of the $W$ propagator.
To overcome this small value, it was 
suggested~\cite{aesmmprl93} that 
the above PRA gets a larger contribution proportional to $\mIm G^W_T 
\times \eRe{\rm (tree \times loops)}$, where 
$G_T$ is the transverse part of the $W$ propagator; indeed this
contribution was found to be $\sim 10^{-5}$, which is still too
small to be of experimental interest.
~\\
\underline{PRA in MSSM:}\\ 
${\rm PRA}(t \to b W)$ is proportional to tree $\times$ loops with
supersymmetric partners in loops. In one limit,~\cite{gkplb93}
a $\tilde t \tilde b \tilde g$-loop is considered.
Such a contribution requires the already excluded condition
$m_t>m_{\tilde t}+m_{\tilde g}$. In another limit,~\cite{cfplb94} a
$\tilde t \tilde b \tilde\chi^0$-loop was suggested. Such a term, which is
proportional to ${\rm arg}(\mu)$,
leads to a PRA $\sim 2\%$, for $m_{{\tilde b}}=100$ GeV.
However this requires 
${\rm arg}(\mu) \sim {\cal O}(1)$, which seems to be excluded by $d_n$.
In another limit it was
assumed~\cite{basprd98} that ${\rm arg}(\mu)=0$ and  
that $b$ squarks are degenerate.
The PRA arising from a $\tilde t \tilde\chi^+\tilde\chi^0$-loop, which is 
proportional to $\xi_{CP}^t$, is at best $\lsim 0.3\%$
for $m_{\tilde g},~m_{\tilde q} \gsim 300$ GeV and 
requires $m_t>m_{\tilde t}+m_{\tilde\chi^0}$.

\subsection{Beyond PRA for $t \to b \tau \nu$ in 3HDM}
There are ways to utilize
${\rm W-''tree''} \times {\rm H-tree}$ and bypass 
the small $G_L^W$ factor
and the CP-CPT connection, to obtain CP-violating asymmetries much
larger than PRA.~\cite{aesmmprl93} To this end one defines, for example, 
the energy asymmetry:

\be
{\cal A}_{E}=
\frac{\langle E_{\tau^+} \rangle - \langle E_{\tau^-} 
\rangle}{\langle E_{\tau^+} \rangle + \langle E_{\tau^-} \rangle}.\label{eq:AE}
\ee

\n It was found that ${\cal A}_E \simeq 10^{-3} >> {\rm PRA}$.

Furthermore, one can define various $\tau$ polarization asymmetries, such as 
the transverve one:~\cite{aesprl93}

\be
{\cal A}_z = 
\frac{\tau^+(\uparrow) - \tau^+(\downarrow) - \tau^- 
(\uparrow) + \tau^- (\downarrow)}{\tau^+
(\uparrow) + \tau^+(\downarrow) + \tau^- 
(\uparrow) + \tau^- (\downarrow)},
\ee

\n where in the $\tau$ rest frame  
${\vec p}_t$ is on the $-x$ axis and $x - y$ is the decay plane.
Since ${\cal A}_z$ is CP-odd and $T_N$-odd, 
there is no need for an absorptive phase. Note that, by avoiding
$\tau$-spin summation, there is no $m_\tau$ suppression in  
${\cal A}_z$. Indeed,
${\cal A}_z \sim 
(m_t/m_\tau) {\cal A}_{E} \sim {\rm few} \times 10^{-1}$ 
requiring $\gsim 10^3$ $t$ quarks, which is a gratifying result. 

\section{CP-Violating Top Dipole Moments}
There are three top dipole moments that may signify the 
presence of CP-violation. They
can be considered as CP-odd form factors in the
$\gamma t \bar t$,
$Zt \bar t$, or the $g t \bar t$ vertices that measure the
effective coupling between the spin of a
short-lived top quark and an external gauge field:

\be
\Gamma_{\mu}^{\gamma,Z,g} = id^{\gamma,Z,g}_t(s) \bar u_t(p_t) 
\sigma_{\mu \nu} \gamma_5 p^{\nu} v_{\bar t}(p_{\bar t});~ p=p_t + p_{\bar t}.
\ee

In the SM $d_t$     
is a 3-loops effect,~\cite{bernrmp91} thus
$d_t^{\gamma} \lsim 10^{-30}$ e-cm, is unmeasurable. We therefore
present below results for extensions of the SM. The numerical values for
$d_t^{\gamma}$,
for typical parameters as described below,  
are displayed in Table~\ref{table:dt}. The corresponding values
for $d_t^{Z}$  are slightly smaller.
~\\
\underline{Neutral Higgs $\cp$ in MHDM:~\cite{sxprl92}}\\ 
 From Eq.~\ref{eq:hff} we see that
$d_t$ goes like $m_t^2 a_t \times b_t$. The results in 
the table are for 
$a_t=b_t=1$, $m_{H^0_i} \sim {\cal O}(1)$ TeV, $i>1$ (note that 
$h \equiv H^0_1$).
~\\
\underline{$H^+$ $\cp$ in MHDM:~\cite{absprd95}}\\
 From Eq.~\ref{eq:Hplusff} we find that
$d_t$ goes like  $m_t m_b \mIm(U_tU_b^*)$. The numbers in Table~\ref{table:dt}
are for $\mIm(U_tU_b^*)=5$, $m_{H^+_i} \sim {\cal O}(1)$ TeV, $i>1$.
~\\
\underline{MSSM $\cp$ from 
${\tilde t}_L - {\tilde t}_R$ mixing:~\cite{bartlnp98}}\\
 From Subsec.~\ref{subsec:cpinmssm}
we see that $d_t \propto \xi_{CP}^t$
which we take to be $0.5$ and 
$m_{\tilde t_{1,2}}=50,400$ GeV.

\begin{table}[htb]
\caption{CP-violating top EDMs, taken at
$500,~1000$ GeV, in three models (for
more details see text). Masses are in GeV.}\label{table:dt}
\begin{center}
\begin{tabular}{|r|r|r|r|} \cline{2-4}
\hline
$d_t^{\gamma}$ & neutral Higgs & 
charged Higgs & Supersymmetry \\
$({\rm e-cm}) \Downarrow$ & 
$m_{h}=100 -300$ & $m_{H^+}=200 -500$ & 
$m_{\tilde g}=200-500$ \\ 
\hline
$500$ GeV & $(4.1-2.0)\times 
10^{-19}$ & $(29.1-2.1)\times 
10^{-22}$ & $(3.3-0.9)\times 10^{-19}$ \\
$|\mIm(d_t^\gamma)|$&  & & \\ 
$1000$ GeV & $(0.9-0.8)\times 
10^{-19}$ &$(15.7-1.0)\times 
10^{-22}$ & $(1.2-0.8)\time 10^{-19}$ \\ 
\hline
$500$ GeV& $(0.3-0.8)\times 
10^{-19}$ & $(33.4-1.5)\times 
10^{-22}$& $(0.3-0.9)\times 10^{-19}$ \\ 
$|\eRe(d_t^\gamma)|$& & & \\ 
$1000$ GeV& $(0.7-0.2)\times 
10^{-19}$  & $(0.3-2.7)\times 
10^{-22}$ & $(1.1-0.3)\times 10^{-19}$ \\ 
\hline
\end{tabular}
\end{center}
\end{table}

\n We observe that all the values in Table~\ref{table:dt} are
at least $7$ orders of magnitude larger than $d_t^{SM}$. Unfortunately,  
as we see in the next section,
only few of the pedictions are (marginally) measurable.\footnote{Also 
note that for MHDM $d_t^g$ in $g_s-{\rm cm} = 
(3/2) d_t^\gamma$ in ${\rm e-cm}$; this
does not hold for MSSM due to an additional $g \tilde g \tilde g$ loop.}

\section{$\cp$ in Future Collider Experiments}
In this section, we discuss $\cpint$ for top pair production,
and for single top production in future experiments.

\subsection{Model Independent Studies of $e^+e^-,~q \bar q \to t \bar t$}
Let $\Sigma$ be a differential cross-section, and $d\phi$ a
phase space element. Then the contribution of $d_t^V,~V=\gamma,~Z$
through $V$ $s$-channel exchange, is added to the SM ($d_t=0$):
 
\bea
\Sigma(\phi)d\phi&  = &\Sigma_0(\phi)d\phi \nonumber \\ 
& + &\sum_{V=\gamma , Z} \left[ \eRe d_t^V(s) 
\Sigma_{\eRe (d_t^V)}(\phi) + \mIm d_t^V(s) 
\Sigma_{\mIm (d_t^V)}(\phi) \right] d\phi, \label{eq:dt1}
\eea

\n for  $e^+e^-,~q \bar q \to t \bar t$ (for $q \bar q$, $g$ exchange
should be added). In Eq.~\ref{eq:dt1}, the usually smaller,
$\cp$ in $t$ decays is neglected.
Note also, that in
MSSM $d_t$ is not the whole story due to new box diagrams.

To minimize the statistical error, optimal observables:

\be
{\cal O}_{\eRe}
=\Sigma_{\eRe (d_t^V)}/\Sigma_0 ~,~ {\cal O}_{\mIm}=
\Sigma_{\mIm (d_t^V)}/\Sigma_0,
\ee

\n were introduced.~\cite{asprd92}
Then $\eRe, \mIm d_t \sim 10^{-17}$ e-cm, can be reached--at a  
$1 \sigma$ level--with 
$10^4$ $t \bar t$ and $\sqrt s=500$ GeV (NLC).
Optimal observables are by now extensively used.

The results improve by considering~\cite{bbozpc96}
$\Sigma(t \bar t) 
\to \Sigma(t \bar t \to b \ell \nu + b W_{\rm had})$ + beam 
polarization. One can then go down to
$\eRe , \mIm d_t \sim {\rm few} 
\times 10^{-19} - 10^{-18}$ e-cm (at $1 \sigma$). 
Many more observables were suggested, but none is doing better
than the above. In view of the models results (see Table~\ref{table:dt}),
this is rather discouraging.

\subsection{\cpinsub in $pp \to t \bar t +X$}
Gluon fusion dominates at the LHC. In a 
seminal work,~\cite{sp92} $\cp$ was studied in
$gg \to t  \bar t \to b W^+ + 
\bar b W^- \to b e^+ \nu_e + \bar b e^- \bar\nu_e$. 
The energy asymmetry
${\cal A}_E$, defined in Eq.~\ref{eq:AE} (with $\tau \to e$), 
requires an absorptive part, which in
a 2HDM with neutral Higgs $\cp$ (with box and triangle diagrams),
is already there 
as $\hat s > 4m_t^2$.
A non-vanishing ${\cal A}_E$ is due to a different 
number of $t_L \bar t_L$ from $t_R \bar t_R$.
${\cal A}_E \sim {\cal O}(10^{-3})$ is possible, requiring
$\gsim 10^7 ~t \bar t$, which is about the number expected at
the LHC. This work was susequently extended ({\it e.g.} to 
MSSM~\cite{splb92}) but 
the results are again at most of
${\cal O}(10^{-3})$. Since the $gg$ luminosity is so much larger than the
$q \bar q$ luminosity at the LHC, the fact that there are more 
quarks than anti-quarks in the proton, cannot fake--through
$q \bar q \to t \bar t$--the CP-violating signal as long as it is
$\geq {\cal O}(10^{-3})$. Furthermore, the effect of QCD can be neglected.
It still remains to be seen whether
detector systematics can be overcome for such a signal.

\subsection{\cpinsub for Single Top Production
in Tevatron Run 3}
The subprocess here is
$u \bar d \to t \bar b \to b W^+ + \bar b$,
in 2HDM~\cite{abesprd96} and MSSM;~\cite{basprd98}
it is obviously irrelevant for the LHC.
For both models $\cp$ is a loop effect, 
with only a triangle contribution for 2HDM
and both triangle and box for MSSM. For the latter case, the box-loops
are negligible. Thus for both models $\cp$ stems 
from the effective $tbW$ production vertex, which in 
turn is much larger than $\cp$ from $t$ decays. 
In the 2HDM the basic vertex from which $\cp$ 
originates is the $i b_t \gamma_5$ part in ${\cal L}_{htt}$
(see Eq.~\ref{eq:hff}), while in MSSM it is generated
from ${\tilde t}_L - {\tilde t}_R$ mixing 
(see Subsec.~\ref{subsec:cpinmssm}).
As we discuss below, $\cp$ in $tb$ production 
can reach a few percents 
($\cp$ in top pair production is at most a few tenths of a percent),
which is good news for TeV3.

Consider the CP-violating top polarization asymmetries:

\be
A_i \equiv \frac{N(\uparrow_i) - 
N(\downarrow_i) + \bar N (\uparrow_i) - \bar N(\downarrow_i)}{N(\uparrow_i) + 
N(\downarrow_i) + \bar N (\uparrow_i) + 
\bar N(\downarrow_i)} ,~~~  i= x,~y,~z,
\ee

\n where the event plane is the $x - z$ plane. 
While $A_0$ (the cross-section asymmetry),
$A_x$ and $A_z$ are $T_N$-even, $A_y$ is $T_N$-odd.
Next, we present numerical results for some sets of
the unknown parameters (for detailed studies, see the original papers).
~\\
\underline{Contribution from 2HDM 
(numerical results):}\\
If $\sqrt s =2$ TeV , $\tan\beta=0.3$, then for $m_h$ from  
approximately $200$ to about $100$ GeV
$A_0 \simeq 1\%$ reaching $\sim 1.5\%$, 
$A_z \simeq 0.9\%$ increasing to about $1.2\%$, 
$A_y \simeq 0.5\%$ decreasing to approximately $0.2\%$ ($A_y$
reaches its maximal value of about $0.6\%$ around
$m_h \sim 400$ GeV) and
$A_x \simeq 0.6\%$ rising to around $0.7\%$.
~\\
\underline{Contribution from MSSM
(numerical results):}\\
Let us quote here some (optimistic) results for
the cross-section asymmtery. 
If $m_{\tilde g}=400$ GeV, $\tan \beta=1.5$, $m_{{\tilde t}_1}=50$ GeV, 
$m_{\tilde q}=m_{{\tilde t}_2}=400$ GeV and $\xi_{CP}=0.5$, then for
$\mu \simeq -70,-90$ GeV, $A_0$ is somewhat larger than $1.5,2.5\%$, 
respectively. 
This is certainly an encouraging result.

\subsection{Tree-level \cpinsub in 
2HDM at an  NLC with ${\protect\sqrt s} \sim 1$ TeV}
A search for tree-level $\cp$ originating from
the $b_t$ term in ${\cal L}_{htt}$ in 2HDM (see Eq.~\ref{eq:hff}),
was suggested for    
$e^+e^- \to t \bar t h$~\cite{baemsprd96}  
and for 
$e^+e^- \to t \bar t Z$~\cite{basplb98}. For both processes $\cp$
is proportional to $b_t \times c$, where 
$c g_{\mu\nu}$ 
is the $hZZ$ coupling, which in our case is a function of
$\tan\beta$ and of the mixing angles in the neutral Higgs sector 
$\alpha_i,~i=1,2,3$.
None can be considered as a Higgs ``discovery'' channel, but 
once it is discovered its couplings can be studied in a clean
environment.
Each process has two types of tree diagrams: 
The first process has 
real $h$ emission from the $Z$ propagator (which goes like $c$)
and from an external $t$ (or $\bar t$) which includes $b_t$.
The second reaction has
real $Z$ emission from the initial and final fermions
and another type of diagram
where $Z$ is emitted from a $ZZh$ vertex 
with the virtual $h$ turning into $t \bar t$.

Consider $A_{\rm opt}=\langle O \rangle / \sqrt{\langle O^2 \rangle}$,  
where $O={\vec p}_{e^-} \cdot \left( {\vec p}_t                       
\times {\vec p}_{\bar t}\right)$.
Due to the large mass of each of the outgoing particles,
one has to go to a next NLC, with $\sqrt s > 800$ GeV,
to obtain significant results. Furthermore, $\tan\beta$ 
has to be of ${\cal O}(1)$. As an example of the numerical results,
we take  $e^+e^- \to t \bar t h$ at $\sqrt s=1$ TeV.
The parameters of the model are taken as
$\tan\beta=0.5$ and 
$\left\{ \alpha_i,~i=1,2,3 \right\}=\left\{\pi/2,\pi/4,0 \right\}$.
Then, when $m_h$ varies between $100$ and $360$ GeV, $A_{\rm opt}$ increases
from approximately $16\%$ to about $27\%$. The 
expected statistical significance of the $\cp$ signal
is $N_{SD}=\sqrt{\sigma {\cal L}} \times A_{\rm opt}$. For the above range of
$m_h$ and for ${\cal L}=200~{\rm fb}^{-1}$, $N_{SD}$ decreases 
from around 2 to 1. For $\sqrt s=1.5$ TeV, ${\cal L}=500~{\rm fb}^{-1}$ 
and the same $m_h$ range, $N_{SD}$ varies between 4 and slightly above 3.

\section{Summary and Outlook}
Before getting into the the summary, note that--for the sake of
imposed brevity--we do not discuss here other interesting
issues in $\cpint$. These include $\cp$ in $h \to t \bar t$,
more coverage of $e+ e^- \to t \bar t$, $\cp$ in $\mu^+ \mu^-$ colliders and 
$\gamma \gamma$ collisions. Hopefully, such omissions
will be rectified elswhere.~\cite{ourreview} 

In summary, we can say that in view of the extremely 
small SM effect, any observation of $\cpint$ will 
indicate the presence of New Physics.
Due to its large mass, the top quark is very sensitive to
beyond the SM scenarios, and it decays so fast that it evades 
hadronic complications.

Finally, in Table~\ref{table:time} we present our outlook
regarding the prospects for observing $\cpint$ in future
accelerators.

\begin{table}[h!]
\caption{An optimistic timetable--of topics discussed here-- 
of $\cpint$ versus time,
where a $\surd~$ stands for ``likely'', $\times$ means ``unlikely''
and $\surd ~\times$ represents an ``in-between'' situation.} \label{table:time}
\begin{center}
\begin{tabular}{|c|c|c|c|c|} 
\hline
\begin{minipage}{10em} \vspace{1.0ex} \begin{center}
$t \to bW$ \\
{(MSSM $\cp$)}
\end{center} \vspace{0.5ex} \end{minipage}
& $\times$ & $\surd$ & $\times$ & \\ \hline  
\begin{minipage}{10em} \vspace{1.0ex} \begin{center}
$t \to b \tau \nu_\tau$ \\
{($H^+$ $\cp$)}
\end{center} \vspace{0.5ex} \end{minipage}
& $\surd$ & $\surd$ & $\surd$ & \\ \hline  
\begin{minipage}{10em} \vspace{1.0ex} \begin{center}
$p \bar p \to t \bar b$ \\
{(MSSM $\cp$)}
\end{center} \vspace{0.5ex} \end{minipage}
& $\surd$ & & & \\ \hline  
\begin{minipage}{10em} \vspace{1.0ex} \begin{center}
$e^+ e^-,~$\raisebox{0.3ex}{$p$}$\stackrel{(-)}{p} \to t \bar t$ \\
{(MSSM $\cp$ \& neutral Higgs $\cp$)}
\end{center} \vspace{0.5ex} \end{minipage}
& $\times$ & $\surd$ & $\surd ~\times$ & \\ \hline  
\begin{minipage}{10em} \vspace{1.0ex} \begin{center}
$e^+ e^- \to t \bar t h,~Z$ \\
{(neutral Higgs $\cp$)}
\end{center} \vspace{0.5ex} \end{minipage}
& & & & $\surd$ \\ \hline  
{} & $\sim 2005$ & $2005$ & $\sim 2010?$ & $>2010?$ \\
{} & TeV3        & LHC    & $500$ GeV    & $\geq 1$ TeV \\ 
{} &             &        & NLC          &NLC \\ \hline  
\end{tabular}
\end{center}
\end{table}

There are several check-marks, but no gold-plated reaction or observable.

\section*{Acknowledgments}
We would like to thank our collaborators D. Atwood, J.L. Hewett,
R. Mendel, R. Migneron and A. Soni. The research of SBS is supported
in part by US DOE contact number DE-FG03-94ER40837. GE acknowledges
partial support from the Israel-USA Binational Science Foundation and
from the VPR Fund at the Technion.

\end{document}